\documentclass[reprint,NumberedRefs]{JASA}
\usepackage[none]{hyphenat}
\usepackage{placeins}
\begin{document}
\title[]{Numerical and perceptual validity of synthetic Head-Related Transfer Functions at scale}

\author{Katarina C. Poole}
\author{Lorenzo Picinali}
\affiliation{Dyson School of Design Engineering, Imperial College London, London SW7 2DB, United Kingdom}
\email{katarina.poole@imperial.ac.uk}
\thanks{Preprint. Submitted to \emph{JASA}; not yet peer reviewed.}
\begin{abstract}
Individually measuring head-related transfer functions (HRTFs) at scale remains a central challenge for personalised spatial audio, motivating growing interest in synthetic HRTFs. We evaluated the numerical, computational, and behavioural validity of synthetic HRTFs, generated through the boundary element method simulation using Mesh2HRTF, against measured and KEMAR HRTFs using the Extended SONICOM dataset. Across 200 subjects, synthetic HRTFs deviated less from measured than KEMAR in interaural time and level differences, but residual errors, together with elevated spectral distortion, concentrated at low, rear elevations. This is consistent with the omission of torso geometry from the synthesis pipeline. Two computational models revealed a corresponding pattern of predicted localisation errors, with synthetic HRTFs positioned between measured and KEMAR. In a virtual reality localisation task (N = 20), synthetic HRTFs matched measured on every polar metric, while KEMAR was significantly worse. However, behavioural error clustered around the front-back midline regardless of condition, not at the low elevations implicated numerically or by the models. A separate spatial release from masking task (N = 18) showed no effect of HRTF type. Together, these results indicate that high-resolution synthetic HRTFs preserve behavioural localisation performance, despite discrepancies between the numerical/model-predicted bias and the spatial pattern of behavioural error.
\end{abstract}


\maketitle

\section{Introduction}
The ability to reproduce spatial sound over headphones is fundamental to a wide range of immersive audio applications. In binaural audio systems, perceptual realism relies on accurately reproducing how sound is shaped by a listener's anatomy before reaching their eardrums, achieved through Head-Related Transfer Functions (HRTFs). These are direction-dependent acoustic filters unique to each individual. HRTFs encode key localisation cues, including interaural time differences (ITDs), interaural level differences (ILDs), and monaural spectral features arising from reflections and diffractions by the pinnae, head and torso. While ITDs and ILDs primarily guide lateral localisation, spectral cues are essential for polar localisation and resolving front-back ambiguities. Critically, these spectral cues are highly individual, small differences in pinnae geometry can produce large differences in the resulting filtering, meaning that HRTFs measured from one person are unlikely to provide accurate spatial cues for another \cite{picinali_system--user_2023}. Consistent with this, studies have shown that using non-individual HRTFs, such as a KEMAR HRTF, significantly degrades spatial perception, leading to increased localisation errors \cite{middlebrooksIndividualDifferencesExternalear1999}, greater front-back confusions \cite{stittAuditoryAccommodationPoorly2019,wenzelLocalizationUsingNonindividualized1993}, and even reduced speech intelligibility \cite{gonzalez-toledoSpatialReleaseMasking2024}, emphasising the importance of individual HRTFs for perceptual accuracy.

Despite their importance, individual HRTF acquisition remains a major bottleneck. Conventional methods require anechoic environments, extensive loudspeaker arrays, and precise microphone placements, making them resource-intensive and difficult to scale \cite{isaacSonicomHRTFDataset2023,majdak3DLocalizationVirtual2010}. Two main research directions have emerged to address this. The first leverages large-scale HRTF datasets to support machine learning approaches for HRTF synthesis and interpolation  \cite{hartungComparisonDifferentMethods1999,porschmannDirectionalEqualizationSparse2019,hoggHRTFUpsamplingGenerative2024}, with numerous public datasets developed over the past two decades including CIPIC, LISTEN, SADIE II, ARI, and SONICOM \cite{algaziCIPICHRTFDatabase2001,carpentierMeasurementHeadRelatedTransfer2014,armstrongPerceptualEvaluationIndividual2018,majdak3DLocalizationVirtual2010,isaacSonicomHRTFDataset2023,pooleExtendedSONICOMHRTF2025a}. The second direction explores synthetic HRTFs generated directly from morphological data via Boundary Element Method (BEM) simulations using tools such as Mesh2HRTF yielding full-resolution individual HRTFs without anechoic recordings \cite{ziegelwangerNumericalCalculationListenerspecific2015,dinakaranPerceptuallyMotivatedAnalysis2018,brinkmannRecentAdvancesOpen2023}. This approach scales readily with 3D scan availability and avoids recording environment variability, making it a promising route to large-scale individual HRTF generation.

Initial numerical evaluation of  synthetic HRTFs has been encouraging. Brinkmann et al. (2019) \cite{brinkmannCrossEvaluatedDatabaseMeasured2019} numerically compared measured and synthetic HRTFs for 96 subjects from the HUTUBS \cite{brinkmann_hutubs_nodate} dataset, and 10 from the SYMARE \cite{zolfaghari_large_2014} dataset, finding generally small spectral deviations below $\approx$ 1 dB up to 1 kHz rising to $\approx$ 7 dB at 17.6 kHz, with a consistent excess of high-frequency energy in the synthetic HRTFs. This surplus, particularly around 8 kHz, was associated with upward perceptual shifts in source position, likely due to the omission of hair and clothing absorption effects \cite{blauertSoundLocalizationMedian1969,katzBoundaryElementMethod2001}.  Spectral similarity is commonly quantified using log-spectral distortion (LSD), which measures frequency-dependent deviations between HRTFs. However, LSD may be overly sensitive to fine-grained spectral variations that are not perceptually relevant \cite{asanoRoleSpectralCues1990, xieAudibilitySpectralDetail2010}, and does not indicate whether deviations occur at perceptually important frequencies or locations. Numerical comparisons therefore identify where synthetic HRTFs deviate from measured, but must be complemented by perceptual evaluation to establish whether those deviations are behaviourally meaningful. 

Using the SAQI test \cite{lindauSpatialAudioQuality2014} with 42 listeners, Brinkmann et al. (2019)\cite{brinkmannCrossEvaluatedDatabaseMeasured2019} found that synthetic HRTFs were perceived as having slightly attenuated lows, emphasised highs, and small spatial offsets ($\approx$ 2° clockwise, $\approx$ 12° upward). A more recent study reported similarly minor localisation errors in the horizontal plane \cite{sunEvaluatingConsistencyNumerical2025}, though limited by coarse loudspeaker spacing (30° intervals). Subjective quality measures such as SAQI and MUSHRA capture abstract perceptual dimensions but are better suited to expert listeners than direct spatial hearing performance. Localisation tests provide a more direct measure of HRTF validity, while spatial release from masking tasks extend evaluation to functional aspects of spatial hearing. Together these approaches provide complementary behavioural benchmarks, but their scalability is limited. Perceptual modelling addresses this by simulating auditory processing at scale. The sagittal-plane model of Baumgartner et al. (2014)\cite{baumgartnerModelingSoundsourceLocalization2014} uses probabilistic template-matching to predict elevation and front-back localisation, while the full-sphere Bayesian model of Barumerli et al. (2023)\cite{barumerliBayesianModelHuman2023} integrates binaural cues with perceptual priors. Both offer scalable, perceptually grounded evaluation across large datasets, yet neither has been applied to synthetic HRTFs at scale.

Despite this initial work, the perceptual validity of synthetic HRTFs remains poorly understood. The studies reviewed above  have been restricted to small samples, numerical evaluations, subjective quality ratings, and coarse horizontal plane localisation, leaving open whether synthetic HRTFs support full-sphere localisation, produce spatially structured localisation errors, or preserve functional spatial hearing beyond localisation. No study has yet combined large-scale numerical analysis with computational modelling and behavioural evaluation to comprehensively assess synthetic HRTF validity. Addressing this is critical for establishing whether BEM-based synthesis represents a viable route to individual HRTFs in practice. The Extended SONICOM dataset \cite{pooleExtendedSONICOMHRTF2025a}, comprising 200 subjects with matched acoustic measurements, high-resolution 3D scans, and corresponding synthetic HRTFs, provides a unique opportunity to address this gap. The present study addresses four research questions: how synthetic HRTFs differ numerically from measured HRTFs relative to KEMAR and whether deviations are spatially localised (\textbf{RQ1}); whether computational models of spatial hearing predict systematic localisation errors for synthetic HRTFs across a large population and where those errors are largest (\textbf{RQ2}); whether synthetic HRTFs preserve behavioural localisation performance and whether spatial error patterns align with numerical deviations (\textbf{RQ3}); and whether synthetic HRTFs preserve functional spatial hearing benefits in a speech-in-noise task (\textbf{RQ4}). 

To address these questions, numerical differences between synthetic, measured, and KEMAR HRTFs were characterised across the full 200-subject dataset using ITD, ILD, and spectral distortion metrics, alongside spatial distribution analyses and computational modelling across the full dataset. Behavioural validation was conducted with 20 participants in a virtual reality localisation task, including spatial error analysis across tested locations, and a further 18 participants completed a spatial release from masking task.

\section{Methods}
\subsection{HRTFs}
Three HRTF conditions were used across both perceptual experiments: individually acoustically measured HRTFs, individually synthesised HRTFs, and a KEMAR mannequin HRTF (large ears). For the numerical analysis, a fourth condition was included consisting of a randomly selected measured HRTF from the SONICOM dataset, serving as a non-individual human baseline. All HRTFs were drawn from the Extended SONICOM dataset (Poole et al., 2025) and sampled at 48 kHz.For the numerical analysis, a fourth condition was included consisting of a randomly selected measured HRTF from the SONICOM dataset, serving as a non-individual human baseline. All HRTFs were drawn from the Extended SONICOM dataset \cite{pooleExtendedSONICOMHRTF2025a} and sampled at 48 kHz.

Measured HRTFs were recorded using in-ear microphones positioned at the entrance of the ear canal across 793 source positions spanning -45° to 225° elevation and 360° azimuth at 5° intervals \cite{isaacSonicomHRTFDataset2023}. Synthetic HRTFs were generated from high-resolution 3D head scans of the same participants using BEM simulations via Mesh2HRTF \cite{ziegelwangerNumericalCalculationListenerspecific2015}. Prior to simulation, meshes were aligned to the Frankfort plane, facial hair digitally removed, truncated below the neck, and ear canals occluded to match microphone placement in the acoustic measurements. Meshes were graded to maintain high resolution around the ipsilateral ear \cite{palmCurvatureadaptiveMeshGrading2021}, and HRTFs were simulated from 0–24 kHz in 150 Hz steps across the same 793 source positions. All HRTFs were preprocessed. ITDs were removed using the threshold-based onset detection method \cite{andreopoulouIdentificationPerceptuallyRelevant2017} and with values stored as metadata. HRTFs were then windowed to 256 samples. No headphone equalisation or free-field compensation was applied to any condition, as neither are assumed to affect direction-dependent spatial cues and localisation accuracy \cite{engel_effect_2019,schonstein_comparison_2008}. Each HRTF set was level-normalised to the response at 0° azimuth, 0° elevation to minimise overall level differences between conditions.

\subsection{Participants}
The numerical assessment was conducted using HRTFs from 200 listeners included in the SONICOM and Extended SONICOM dataset \cite{isaacSonicomHRTFDataset2023,pooleExtendedSONICOMHRTF2025a}. Of these, 51\% (n = 102) voluntarily provided demographic information; for this subset, the mean age was 32 years (SD = 9.6), with 82 identifying as male. Participants for the perceptual experiments were drawn from the same dataset, having previously undergone acoustic HRTF measurement. The localisation experiment included 20 paid participants (14 male, mean age = 29 years, SD = 5.8), and the spatial release from masking experiment included 18 paid participants (16 male, mean age = 25 years, SD = 5.5). All participants reported normal hearing and no neurological disorders. 

\subsection{Stimuli}
For the location experiment, the stimuli consisted of three consecutive 100 ms Gaussian noise bursts, each shaped by a Hann window, producing a total stimulus duration of 300 ms. This approach has been shown to improve localisation performance relative to continuous noise of equal duration \cite{maceReachingSoundAccuracy2012}.  Stimuli were generated offline and convolved in real time with the selected HRTF for each trial and condition using a custom VR application developed in Unity, incorporating the 3DTI Tune-In Toolkit via its Unity wrapper \cite{cuevas-rodriguez3DTuneInToolkit2019} for binaural rendering. In the main test, source positions were distributed across 33 distinct directions on the sphere, covering azimuths from -180° to 180° and elevations from -30° to 90°. During the training phase, the same stimulus was presented acoustically through the loudspeaker dome across 29 positions, which did not fully overlap with the 33 main test positions.

For the spatial release from masking experiment, overlapping speech stimuli were drawn from the Coordinate Response Measure corpus \cite{boliaSpeechCorpusMultitalker2000b}, which consists of sentences following the structure ''Ready [callsign] go to [colour] [number] now''. To preserve the high-frequency spectral content required for elevation perception, the original CRM recordings were used, and sentences were bandpass filtered between 200 Hz and 18 kHz \cite{gonzalez-toledoSpatialReleaseMasking2024,martinSpatialReleaseSpeechonspeech2012}. All stimuli were normalised to equal root mean square (RMS) amplitude, such that the target and each masker were presented at the same level, yielding a target-to-masker ratio of approximately -3 dB. Three source positions along the median plane were used: 0°, +50°, and -50° elevation (all at 0° azimuth). Stimuli were rendered offline using the Python wrapper of the 3DTI Tune-In Toolkit (\textit{py3dti}\cite{pauwelsPy3dti2023}), with the same three HRTF conditions as the localisation experiment.

\subsection{Apparatus}

The localisation test took place in the semi-anechoic chamber used for the original HRTF measurements. Stimuli were presented over Sennheiser HD 599SE headphones, with the Meta Quest 2 VR headset providing the visual display. Sagittal, horizontal, and coronal planes rendered as coloured lines on a sphere surrounding the listener, against a starry night background providing minimal visual orientation. Participants responded using a virtual laser pointer controlled via the Quest handheld controllers. Free-field training stimuli were presented over one of two loudspeaker arrays, mounted on a dome around the listener. The first 9 participants trained using a 31-loudspeaker array (Genelec), and the remaining 11 used a 29-loudspeaker array (Wilmslow Audio Ltd, UK; full-range 3-inch Peerless 830987 drivers, with Triad TS-PAMP8-100 class-D amplifiers). Free-field playback was controlled via Open Sound Control (OSC) messages sent from the VR headset to a Max 8 patch, which routed stimulus presentation to the loudspeakers through MOTU 24Ao audio interfaces. A full schematic of the apparatus and loudspeaker array layout is provided in Pirard et al. (2026)\cite{pirardEvaluationHeadRelatedTransfer2026}. The speech test took place in a quiet, non-anechoic room. Stimuli were presented over Sennheiser HD 650 headphones, with participants responding via a graphical interface on a laptop. Overall sound level for both experiments was fixed per participant at approximately 65 dBA, varying slightly with stimulus and position.

\subsection{Procedure}

For the localisation test, each session comprised four blocks of approximately 10 minutes each ($\approx$ 45 minutes total), with short breaks permitted between blocks: a free-field training block followed by three test blocks. At the start of each session, the participant's head was aligned to the loudspeaker array centre, and the VR environment adjusted to match via a passthrough window. At the start of each trial, participants oriented to the front (azimuth = 0°, elevation = 0°), guided by a head-tracked reticule and target disc that turned from red to green once aligned. One second after alignment, the stimulus was presented. Movement during presentation aborted the trial, which was then repeated after realignment. Participants then rotated freely to indicate the perceived source location, confirmed their response via the controller trigger, and realigned to front for the next trial.

During training, responses within 5° of the target were marked correct with a green sphere; larger errors produced a directional arrow colour-graded from yellow to red across 5–35°, followed by up to two correction trials. The target location was shown visually for the first third of training and removed thereafter, requiring participants to localise using free-field acoustic cues alone. Following training, participants completed three test blocks of 99 trials each (297 total), with the 33 source positions and three HRTF conditions fully interleaved and balanced across blocks, yielding three repetitions of each position-condition pairing. No feedback was given during test blocks.

For the speech test, training comprised 18 trials of increasing difficulty, beginning with the target talker alone before maskers were introduced and the target-masker separation progressively reduced. Visual feedback was given via a GUI. The correct colour-number combination was highlighted in green, and, if incorrect, the participant's selection in red. Following training, participants completed three blocks of approximately 11 minutes each (180 trials per block, 540 total), with no feedback. Target location was fixed within a block and counterbalanced across blocks. Within each block, HRTF condition, talker gender, and masker location were fully crossed and randomised, giving 10 repetitions of each combination. On each trial, participants identified the colour (white, red, blue, or green) and number (1–8, excluding 7 as disyllabic) spoken by the target talker, giving a chance performance level of 3.6\%.

\subsection{Analysis}

All acoustic metrics were calculated using the Spatial Audio Metrics toolbox \cite{pooleExtendedSONICOMHRTF2025a} as the difference between each comparison HRTF (synthetic, KEMAR, or randomly selected) and the listener's measured HRTF. ITDs were estimated via threshold-based onset detection: each HRIR was low-pass filtered at 3 kHz (10th-order Butterworth), with onset defined as the point exceeding -10 dB relative to peak amplitude, and ITD taken as the difference between left- and right-ear onsets, positive values indicating the left ear leading \cite{andreopoulouIdentificationPerceptuallyRelevant2017}. ILDs were computed as the RMS amplitude difference between left and right HRIRs at each source position. LSD was calculated as the RMS log ratio between two transfer functions across frequencies (50–18000 Hz), averaged across locations and ears, and computed for both HRTFs and DTFs, the latter derived by removing the common transfer component as implemented in the Auditory Modelling Toolbox \cite{majdakpiotrAMT1xToolbox2022} and transcribed to Python. Unless otherwise stated, all metrics were computed relative to each listener's measured HRTF; comparisons against measured DTFs are noted explicitly in the results.

Two computational models of spatial hearing were applied to the full 200-subject dataset. The Baumgartner et al. (2014) \cite{baumgartnerModelingSoundsourceLocalization2014} model, was applied with notch-region spectral weighting and group-level average perceptually fitted parameters from Lladó et al. (2025\cite{lladoSpectralWeightingMonaural2025}; $\Gamma$ = 3.87, S = -6.77, $\epsilon$ = 13.11°) at a lateral angle of 0°. The Barumerli et al. (2023)\cite{barumerliBayesianModelHuman2023} was applied with parameters from Daugintis et al. (2023\cite{daugintisClassifyingNonIndividualHeadRelated2023}; $\sigma_{ild}$ = 0.75 dB, $\sigma_{mon}$ = 4.3°, $\sigma_{motor}$ = 13.45° and $\sigma_{prior}$ = 11.5°), with 50 iterations per subject and condition. Behavioural localisation responses were converted to interaural coordinates and polar angles were weighted to account for pole compression ($w = 0.5\cdot\cos(2\phi) + 0.5$, where $\phi$ is the lateral target angle). Great circle errors were calculated from the spherical coordinates of target and response locations. Responses were classified into confusion types\cite{poirier-quinotHRTFPerformanceEvaluation2022} and quadrant errors were computed for responses within ±30° lateral angle. Training block trials were excluded from all analyses.

Differences between HRTF conditions in the numerical analysis were assessed using one-way repeated measures ANOVAs with Bonferroni correction and Tukey-HSD post-hoc comparisons, or Friedman tests where normality was violated. Cluster-based permutation tests were applied across azimuth-elevation pairs and frequency bins (\textit{MNE Python\cite{gramfortMEGEEGData2013}}; 1024 permutations, $p$ < 0.05) to identify spatially or spectrally localised differences. Spatial extent was quantified as the proportion of locations in significant clusters. For behavioural localisation errors, a location-shuffling permutation test with family-wise error correction (1000 iterations) was applied per metric and HRTF condition. Locations where observed errors exceeded the 97.5th or fell below the 2.5th percentile of the null distribution were considered significant. Auditory model validity was assessed using mixed effects models predicting observed minus model-predicted performance, with HRTF condition and auditory model as fixed effects and subject as a random effect. For the spatial release from masking experiment, a generalised linear mixed model with a binomial link function was fitted with HRTF type, masker-target elevation separation, and talker gender as fixed effects and subject as a random effect, with Bonferroni-corrected post-hoc comparisons.

\section{Results}
\subsection{Synthetic HRTFs more closely approximate measured HRTFs than KEMAR across binaural and spectral cues}
To establish the degree to which synthetic HRTFs capture individual acoustic features and where systematic deviations arise, all metrics were computed relative to each listener's measured HRTF across 200-subjects, comparing three conditions: the listener's own synthetic HRTF, a KEMAR  HRTF, and a randomly selected human HRTF as a non-individual alternative.

\begin{figure*}
\includegraphics[width=0.85\textwidth]{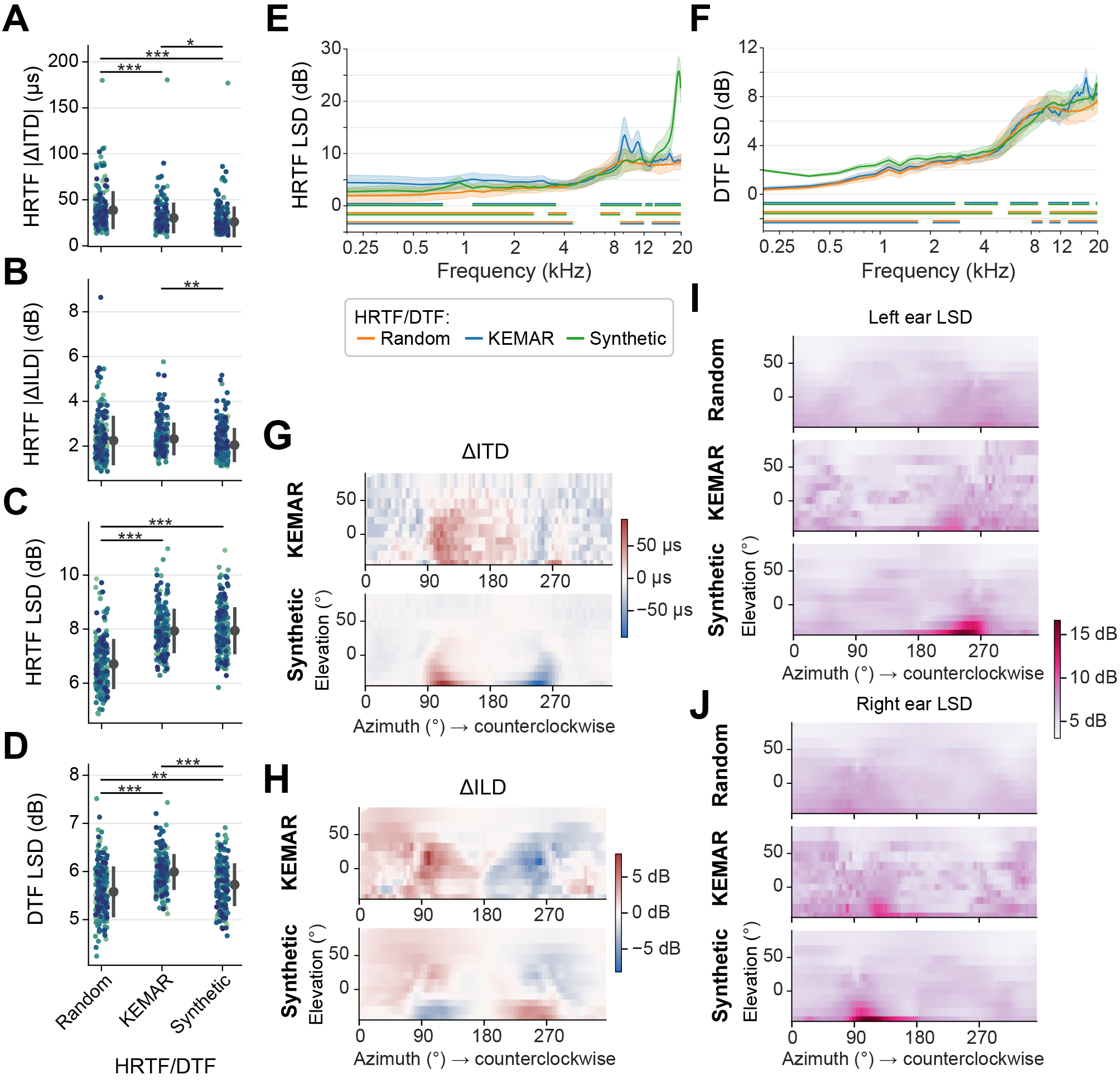}
\caption{\label{fig:fig1}{\textbf{Numerical differences between HRTFs}. Absolute ITD (\textbf{A}) and ILD (\textbf{B}) difference between each HRTF and measured HRTF, averaged across spatial locations. (\textbf{C}) LSD between measured and other HRTFs, averaged across spatial locations and frequencies. (\textbf{D}) LSD computed using DTFs rather than HRTFs. Each coloured dot represents one subject (N = 200) and errors bars indicate the mean and standard deviation. Significant post-hoc pairwise comparisons are indicated with significance bars; *$p < 0.05$, **$p < 0.01$, ***$p < 0.001$. Frequency-dependent LSD for HRTFs (\textbf{E}) and DTFs (\textbf{F}). Shaded regions indicate the standard deviation and coloured paired lines indicate frequency regions showing significant differences ($p < 0.05$) between conditions identified using cluster-based permutation testing. Spatial distribution of signed ITD (\textbf{G}) and ILD (\textbf{H}) differences (comparison HRTF minus measured) as a function of azimuth and elevation; positive (red) and negative (blue) differences. Spatial distribution of LSD for the left (\textbf{I}) and right (\textbf{J}) ear as a function of source location, with darker shading indicating larger distortion values.}}
\end{figure*}

Averaged across spatial locations and subjects, synthetic HRTFs reproduced interaural cues most accurately (see Fig. \ref{fig:fig1}A-B). For ITDs, a significant main effect of HRTF type was observed ($F(2, 398) = 96.89, p < 0.001$), with synthetic HRTFs showing the smallest absolute deviation from measured (26.0 µs), followed by KEMAR (30.3 µs) and random (38.8 µs). ILD differences followed a similar pattern ($F(2, 398) = 11.23, p < 0.001$), with synthetic HRTFs (2.05 dB) deviating significantly less than KEMAR (2.32 dB). For spectral cues (see Fig. \ref{fig:fig1}C-F), synthetic HRTFs maintained low spectral distortion below 6 kHz but showed a marked rise above 16 kHz, while KEMAR exhibited greater distortion in the perceptually salient 8-12 kHz range. Averaged across frequency, the random HRTF showed unexpectedly low LSD. Synthetic DTFs showed significantly lower distortion than the KEMAR DTF, though the random DTFs still showed the lowest overall LSD.

Critically, these differences were not uniformly distributed across space. The random HRTF showed no significant signed ITD or ILD differences relative to the acoustically measured HRTF at any spatial location and is therefore not displayed. For ITDs (see Fig. \ref{fig:fig1}G) cluster-based permutation tests revealed that KEMAR differences were significant across 71.59\% of tested spatial locations, with peak deviations of +68.75 µs at (Az: 105°, El: -20°) and -42.92 µs at (Az: 75°, El: 20°), with a predominantly positive bias concentrated around the left lateral region. Synthetic HRTF differences were more spatially concentrated, reaching significance at 59.85\% of locations, but with larger peak values of +89.58 µs at (Az: 110°, El: -45°) and -91.04 µs at (Az: 250°, El: -45°). This indicates a systematic overestimation of ITD magnitude in the synthetic condition, concentrated at low elevations, whereas the KEMAR has a more spatially variable and asymmetric pattern.

For ILDs (see Fig. \ref{fig:fig1}H), the spatial extent was reversed. Synthetic HRTF differences reached significance across a marginally greater proportion of locations than KEMAR (79.29\% vs 77.27\%), yet KEMAR showed substantially larger peak deviations near the horizontal plane (+8.19 dB at Az: 100°, El: 20°; -7.95 dB at Az: 260°, El: 10°), reflecting an overestimation of ILD magnitude in this region. In contrast, synthetic errors were smaller in magnitude and concentrated behind and below the listener (+4.77 dB at Az: 265°, El: -45°; -4.42 dB at Az: 95°, El: -45°), reflecting a systematic underestimation of ILD magnitude at low elevations. Spectral deviations similarly clustered in the lower rear quadrant across all conditions, with synthetic HRTFs showing the largest local LSD of any condition at -45° elevation (left ear: 15.95 dB; right ear: 16.64 dB). Together, these findings indicate that while synthetic HRTFs largely preserve listener-specific binaural cues, their remaining errors are dominated by low-elevation rear-field spectral inaccuracies, whereas KEMAR deviations are broader and primarily reflect non-individual lateral cue mismatches.

\subsection{Auditory models predict systematic localisation errors for non-individual and synthetic HRTFs}
To assess how HRTF differences might translate into localisation errors at scale, two computational models of spatial hearing were applied to the full 200-subject dataset: the Baumgartner et al. (2014) sagittal-plane model, modified with notch-region spectral weighting and perceptually fitted parameters \cite{lladoSpectralWeightingMonaural2025} and the Barumerli et al. (2023) full-sphere model with perceptually fitted parameters \cite{daugintisClassifyingNonIndividualHeadRelated2023}.

\begin{figure*}
\includegraphics[width=0.85\textwidth]{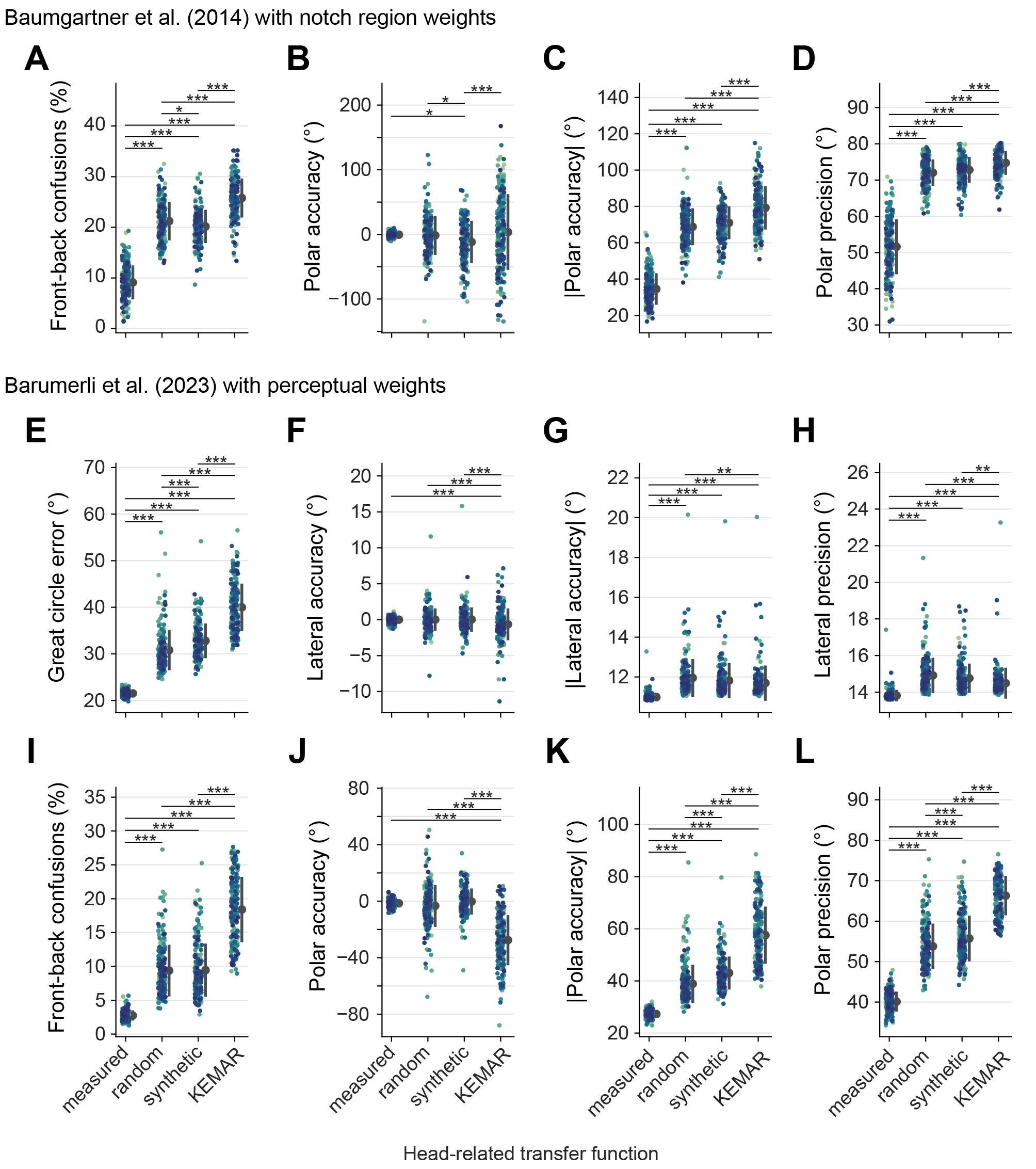}
\caption{\label{fig:fig2}{\textbf{Predicted localisation performance across HRTF types from two computational models of spatial hearing}. (\textbf{A-D}) Baumgartner et al. (2014) sagittal-plane model and the (\textbf{E-L}) Barumerli et al. (2023) full-sphere model. Panels show: front-back confusion rate (\textbf{A, I}), signed polar accuracy (\textbf{B, J}), absolute polar error (C\textbf{, K}), polar precision (\textbf{D, L}), great circle error (\textbf{E}), signed lateral accuracy (\textbf{F}), absolute lateral accuracy (\textbf{G}), and lateral precision (\textbf{H}). All metrics are shown for four HRTF conditions: measured, randomly selected measured, KEMAR and synthetic. Each coloured dot represents one participant (n = 200). Error bars show the mean ± standard deviation. Significant post-hoc pairwise comparisons are indicated with significance bars; *$p < 0.05$, **$p < 0.01$, ***$p < 0.001$.}}
\end{figure*}

Both models produced significant main effects of HRTF type across all metrics (all $p < 0.001$, Bonferroni corrected; see Fig. \ref{fig:fig2}), with a consistent ordering: measured HRTFs yielded the best predicted performance, KEMAR the worst, and synthetic and random HRTFs intermediate. Across polar metrics, measured HRTFs showed the lowest predicted front-back confusion (Baumgartner: 9.11\%; Barumerli: 2.75\%), absolute polar error (34.50°; 27.28°), and polar precision (51.59°; 40.11°), while KEMAR produced the largest errors (front-back confusion: 25.29\%; 18.40\%; absolute polar error: 79.23°; 57.56°; polar precision: 74.72°; 66.31°). Synthetic and random HRTFs occupied intermediate positions across both models, though their relative ordering differed. The Baumgartner model predicted slightly fewer front-back confusions for synthetic than random (20.16\% vs 21.22\%, $p = 0.017$), whereas Barumerli predicted worse absolute polar error for synthetic (43.03° vs 38.88°, $p < 0.001$) and worse polar precision (55.72° vs 53.77°, $p < 0.001$), with front-back confusion indistinguishable between the two (9.45\% vs 9.38\%, $p = 0.998$). Both models agreed that random HRTFs showed less polar bias than synthetic (Baumgartner: -1.52° vs -11.71°, p = 0.029; Barumerli: -3.31° vs -0.24°). 

The lateral dimension, accessible via the Barumerli model, revealed a distinct pattern. Signed lateral accuracy was near zero for measured, random, and synthetic conditions (0.002°, 0.010°, 0.026° respectively), but KEMAR showed a pronounced negative bias (-0.646°, $p < 0.001$ vs all others). For absolute lateral accuracy and precision, all non-individual conditions were significantly worse than measured (all $p < 0.001$) but did not differ from each other. Together, both models predict that individualisation level shapes polar but not lateral localisation, with the largest errors for KEMAR and intermediate errors for synthetic and random HRTFs. Crucially, both models predict that measured HRTFs should yield significantly better localisation performance than synthetic HRTFs across all polar metrics. 

\subsection{Synthetic HRTFs partially preserve localisation performance relative to measured HRTFs}
\begin{figure*}
\includegraphics[width=0.85\textwidth]{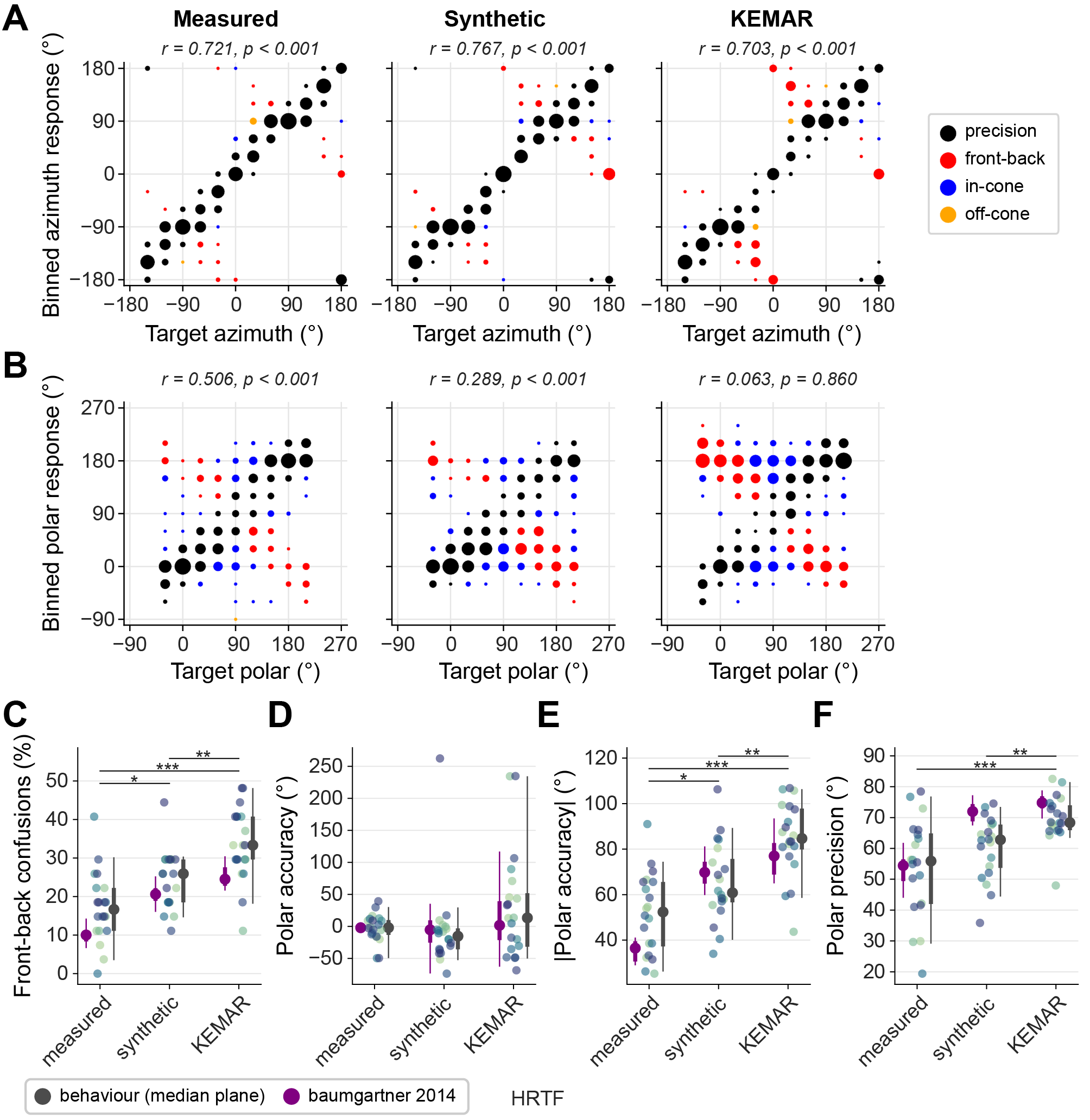}
\caption{\label{fig:fig3}{\textbf{Behavioural localisation performance and Baumgartner model predictions for median plane targets}. Response-target correlations for azimuth (\textbf{A}) and elevation (\textbf{B}), with responses binned to 30° intervals and coloured by error type. Correlation $r$ and $p$ values are annotated on each panel. Localisation metrics restricted to median plane targets (azimuth = 0° and 180°) are shown for front-back confusion rate (\textbf{C}), signed polar accuracy (\textbf{D}), absolute polar accuracy (\textbf{E}), and polar precision (\textbf{F}). In \textbf{C–F}, grey error bars show mean ± standard deviation of behavioural data and purple error bars show Baumgartner model predictions; individual data points are shown coloured by participant. Significant post-hoc pairwise comparisons for behaviour between HRTF conditions are indicated: *$p < 0.05$, **$p < 0.01$, ***$p < 0.001$.}}
\end{figure*}

Behavioural performance in the VR localisation experiment showed a clear azimuth-elevation dissociation across the three HRTF conditions, evident in response-target correlations (see Fig. \ref{fig:fig3}A-B). Azimuth correlations were high for all three HRTF types (measured: $r$ = 0.721, synthetic: $r$ = 0.767, KEMAR: $r$ = 0.703; all $p < 0.001$), whereas elevation correlations along the median plane differed markedly, with measured HRTFs yielding moderate tracking ($r$ = 0.506, $p < 0.001$), synthetic HRTFs reduced but above-chance elevation tracking ($r$ = 0.289, $p < 0.001$), and KEMAR near-zero elevation localisation ($r$ = 0.063, $p = 0.860$). This dissociation was consistent across polar metrics restricted to median plane targets, shown alongside Baumgartner model predictions in Fig. \ref{fig:fig3}C-F.

\begin{figure}
\includegraphics[width=0.45\textwidth]{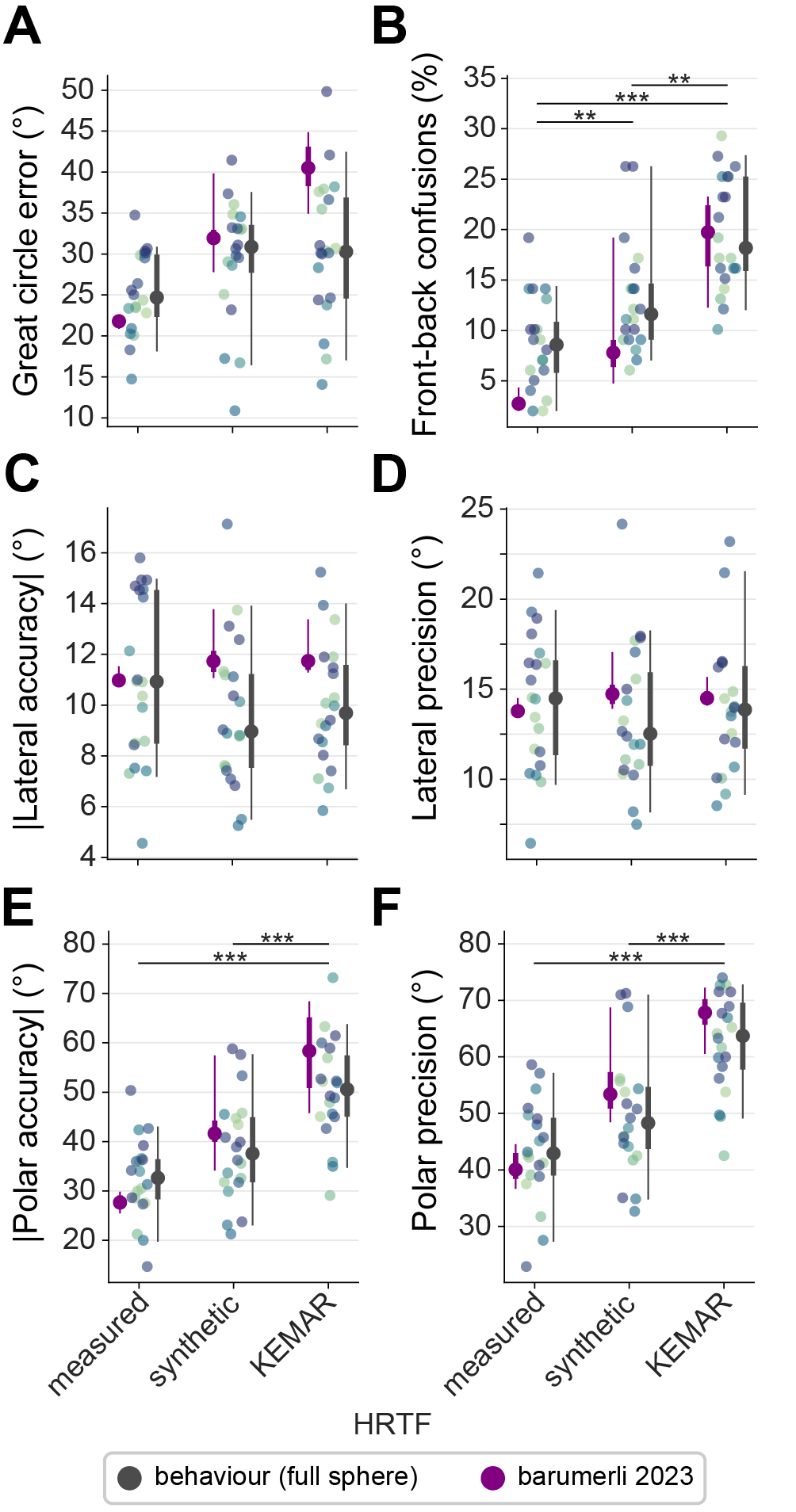}
\caption{\label{fig:fig4}{\textbf{Behavioural localisation performance and Barumerli model predictions for full-sphere targets}. Metrics are shown for great circle error (\textbf{A}), front-back confusion rate (\textbf{B}), absolute lateral accuracy (\textbf{C}), lateral precision (\textbf{D}), absolute polar accuracy (\textbf{E}), and polar precision (\textbf{F}). Grey error bars show mean ± standard deviation of behavioural data and purple error bars show Barumerli model predictions; individual data points are coloured by participant. Significant post-hoc pairwise comparisons for behaviour between HRTF conditions are indicated: *$p < 0.05$, **$p < 0.01$, ***$p < 0.001$.}}
\end{figure}

Full-sphere localisation metrics confirmed this pattern (see Fig. \ref{fig:fig4}). Significant main effects of HRTF were found for polar absolute accuracy ($F(2,38) = 31.4, p < 0.001$), polar precision ($F(2,38) = 29.28, p < 0.001$), and front-back confusion ($Fr = 25.79, df = 2, p < 0.001$). In each case KEMAR produced the worst performance (polar absolute accuracy: 50.56°, polar precision: 59.21°, front-back confusion: 18.18\%), significantly worse than both measured (32.64°, 43.81°, 8.59\%; all $p < 0.001$) and synthetic (37.57°, 46.52°, 11.62\%; all $p < 0.001$), while measured and synthetic did not significantly differ from one another on any polar metric (all $p > 0.1$). The great circle error showed a significant main effect ($F(2,38) = 6.0, p = 0.005$) but no significant pairwise differences after correction (all $p > 0.05$). In contrast, lateral errors were low in absolute terms across all conditions and largely insensitive to HRTF type. Lateral absolute accuracy ranged from 8.96° to 10.93° across conditions, lateral precision showed no main effect ($F(2,38) = 2.24, p = 0.121$), and lateral absolute accuracy showed only a marginal main effect ($F(2,38) = 3.32, p = 0.047$) with no significant pairwise differences. Together these results indicate that differences in the HRTFs selectively shape polar but not lateral localisation, with the largest errors for KEMAR and intermediate errors for synthetic HRTFs. 

To assess model validity, mixed effects models were fitted predicting observed minus model-predicted performance for polar absolute accuracy, polar precision, and front-back confusion (see Supplementary Tables 1-3 for full models). Bias was generally largest for measured HRTFs and significantly smaller for KEMAR across all three metrics (all $p < 0.05$), with synthetic HRTFs showing a similar but weaker pattern, significant only for polar precision. Where the two models could be directly compared, Baumgartner showed significantly greater bias than Barumerli for polar absolute accuracy ($\beta = 11.02, p = 0.012$).

\subsection{Synthetic and KEMAR HRTFs show opposing localisation error spatial patterns}

\begin{figure*}
\includegraphics[width=0.85\textwidth]{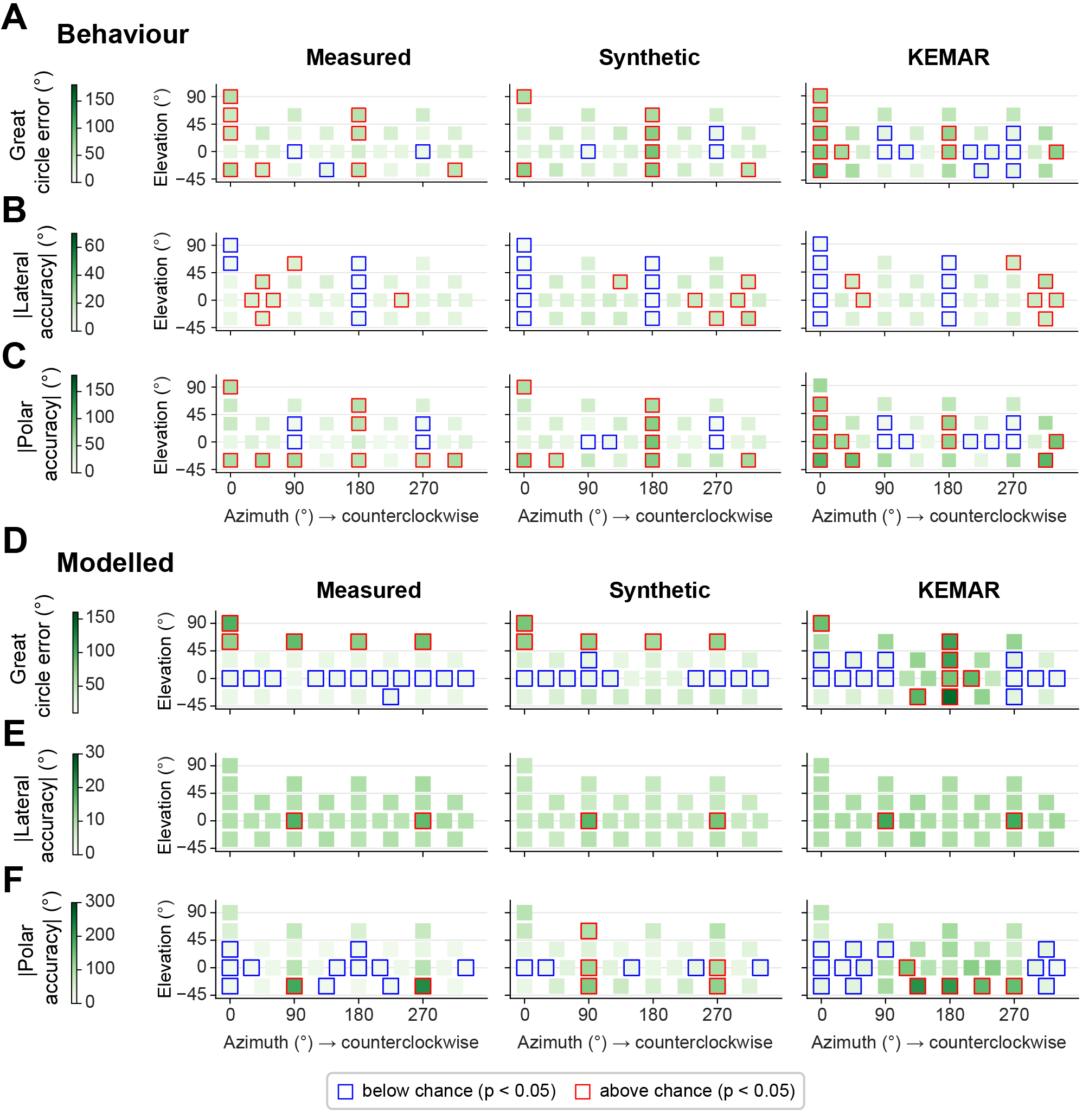}
\caption{\label{fig:fig5}{\textbf{Spatial distribution of localisation errors across HRTF conditions}. Permutation test results for great circle error (\textbf{A, D}), absolute lateral accuracy (\textbf{B, E}), and absolute polar accuracy (\textbf{C, F}) shown as a function of azimuth and elevation for measured, synthetic, and KEMAR HRTFs. (\textbf{A–C}) Observed behavioural errors (N = 20). (\textbf{D–F}) Barumerli (2023) model predicted errors for the same 20 subjects at the same 33 tested locations. Green shading indicates mean error magnitude (darker = larger error). Positions where errors were significantly higher than chance are outlined in red, and positions where errors were significantly lower than chance are outlined in blue (permutation test, $p < 0.05$).}}
\end{figure*}

To examine whether localisation errors were spatially structured, permutation tests identified positions with significantly higher or lower error than chance for each metric and condition. For great circle error, measured and KEMAR errors clustered frontally (0°) while synthetic clustered posteriorly (180°), with fewer errors laterally (90°/270°) across all conditions; KEMAR showed the widest spread (54.5\% of locations) versus measured (36.4\%) and synthetic (30.3\%;  see Fig. \ref{fig:fig5}A). For absolute lateral accuracy, all conditions showed reduced errors frontally and posteriorly, with the highest errors off-axis ($\approx$ 5°, $\approx$ 320°), and a similar proportion of significant locations across conditions (measured: 36.4\%, synthetic: 45.5\%, KEMAR: 48.5\%). For absolute polar accuracy, errors were most common at -30° elevation and minimised laterally: KEMAR clustered frontally across multiple elevations with few posterior errors, synthetic was predominantly posterior, and measured showed a similar but less concentrated pattern, with KEMAR again the most widespread (51.5\%) versus synthetic (36.4\%) and measured (39.4\%).

The same permutation approach was applied to Barumerli (2023) model's predicted errors for the same 20 subjects (see Fig. \ref{fig:fig5}B). For great circle error, measured and synthetic showed above-chance error at high elevations across azimuths, while KEMAR's above-chance error clustered around 180°; synthetic showed the fewest significant locations (33.3\%), compared to measured (51.5\%) and KEMAR (60.6\%). Lateral accuracy showed an identical pattern across conditions, confined to 90°/270° at 0° elevation (6.1\% each). For polar accuracy, measured and synthetic both showed above-chance error at the lateral positions (90°/270°), with synthetic extending across a wider elevation range, while KEMAR's above-chance error spanned low elevations continuously from 90° through 180° to 270°. KEMAR again showed the most widespread significant locations (51.5\%), compared to measured (39.4\%) and synthetic (30.3\%).

To formally assess model validity, spatial correlations were computed between model-predicted and observed error across the 33 locations and 20 subjects. For measured HRTFs, significant positive correlations were found for the great circle error ($r = 0.611, p < 0.001$) and polar accuracy ($r = 0.554, p = 0.001$), consistent with the model's use of individually measured HRTFs as internal templates. For non-individual HRTFs, the model failed to capture spatial error structure: synthetic showed no significant correlation for any metric (great circle error: $r = 0.295, p = 0.096$; lateral: $r = 0.021, p = 0.909$; polar: $r = -0.012, p = 0.948$), and KEMAR showed a significant negative correlation for polar accuracy ($r = -0.470, p = 0.006$), indicating the model predicts high error where behaviour shows low error and vice versa. Lateral accuracy showed no significant correlation for any condition. Together, these results indicate the Barumerli (2023) model captures aggregate localisation errors but not their spatial structure.

\subsection{HRTF type does not significantly modulate spatial release from masking}

\begin{figure}
\includegraphics[width=\columnwidth]{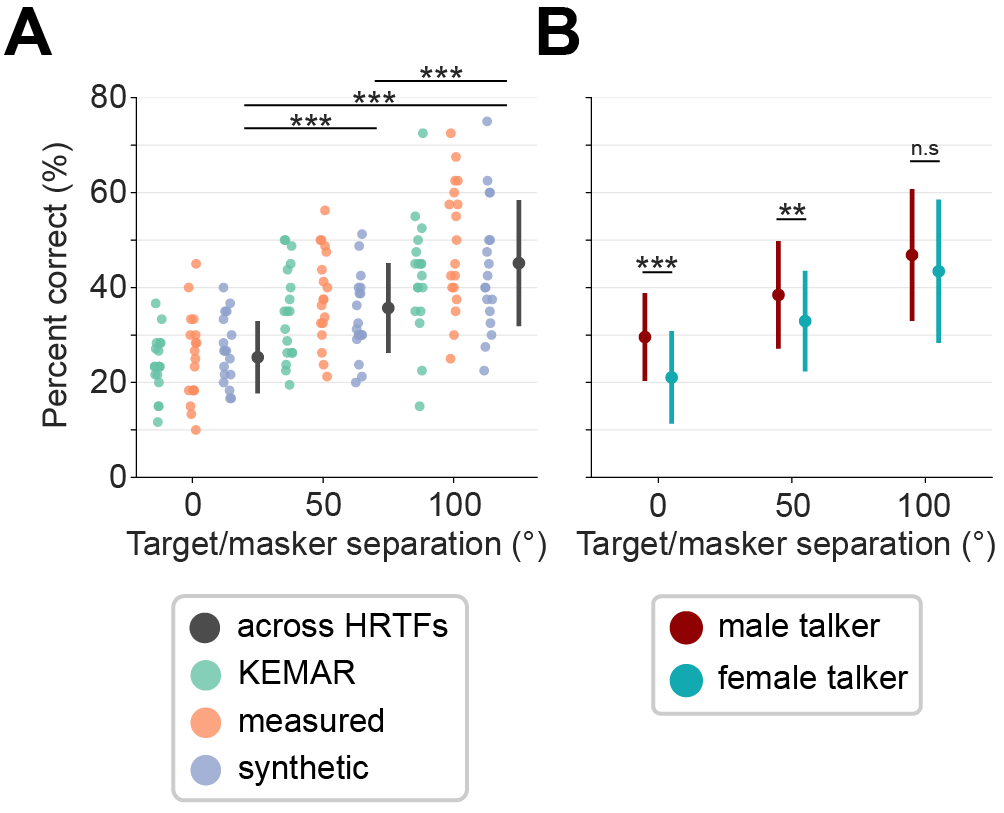}
\caption{\label{fig:fig6}{\textbf{Speech detection performance as a function of masker-target elevation separation and HRTF}. (\textbf{A}) Proportion correct as a function of separation distance, averaged across HRTF conditions (grey error bars) with individual participant means shown as individual points coloured by HRTF (measured: orange, synthetic: purple, KEMAR: green). (\textbf{B}) Proportion correct as a function of separation distance shown separately for female (blue) and male (red) talkers. Error bars show mean ± standard deviation. Significant post-hoc pairwise comparisons between separation distances and talker are indicated: *$p < 0.05$, **$p < 0.01$, ***$p < 0.001$.}}
\end{figure}

To assess whether the localisation errors of non-individual and synthetic HRTFs extend to functional spatial hearing, speech detection performance was measured in a spatial release from masking task, in which participants detected a target talker in the presence of two spatially separated masking talkers rendered at varying elevation separations (N = 18). A generalised linear mixed model (see Supplementary Table 4 for the full model) revealed a significant effect of masker-target elevation separation ($\beta = 0.644$ per 50°, $p < 0.001$), with performance increasing monotonically from 25.3\% correct at 0° separation to 35.7\% at 50° and 45.1\% at 100° (see Fig. \ref{fig:fig6}A). Post-hoc pairwise comparisons confirmed that all three separation distances differed significantly from one another (all $p < 0.001$, Bonferroni corrected), demonstrating a graded spatial release from masking effect under headphone-based rendering. Male talkers were detected more accurately than female talkers overall (38.3\% vs 32.5\%; $\beta = 0.491, p < 0.001$), though this difference was significant only at 0° and 50° separation (both $p < 0.001$) and not at 100° ($p = 0.104$; see Fig. \ref{fig:fig6}B). No significant main effect of HRTF type was found (KEMAR: $p = 0.878$, synthetic: $p = 0.809$), and neither KEMAR (33.6\%) nor synthetic (35.0\%) differed significantly from individually measured HRTFs (37.5\%). No significant interactions between HRTF type and masker-target separation were observed (all $p > 0.08$), indicating that the spatial release from masking benefit did not vary with HRTF individualisation level.

\section{Discussion}
In summary, synthetic HRTFs more closely approximated measured HRTFs numerically than KEMAR across interaural cues, though both showed spatially structured spectral deviations concentrated at low elevations and posteriorly. Both computational models showed a similar ordering of predicted localisation performance, with synthetic HRTFs intermediate between measured and KEMAR for polar metrics and lateral localisation largely preserved. Behavioural testing broadly supported this where synthetic HRTFs produced polar localisation performance indistinguishable from measured, whilst KEMAR was significantly worse. However, the spatial pattern of these numerical and model-predicted deviations was not reflected in the behavioural error, which instead clustered around the front-back midline across all three conditions, indicating factors beyond the numerical and model metrics used here shape where localisation errors occur.

\subsection{RQ1: How do synthetic HRTFs differ numerically from measured HRTFs}

Across the 200 subject sample, synthetic HRTFs preserved binaural and spectral cues more closely than KEMAR, with residual errors concentrated at low, rear elevations rather than evenly distributed, in contrast to Brinkmann et al. (2019)\cite{brinkmannCrossEvaluatedDatabaseMeasured2019}, who reported ITD deviations of similar order but only assessed the horizontal plane. Both deviations exceeded typical discrimination thresholds, given ITD JNDs as low as 20 µs and ILD JNDs of around 0.5–1 dB\cite{klumppMeasurementsInterauralTime1956,millsLateralizationHighFrequencyTones1960}, indicating these residual errors are large enough to be perceptually detectable. This spatial pattern extended to the spectral domain where synthetic HRTFs also showed excess high-frequency energy consistent with Brinkmann et al. (2019) and confirmed here to generalise at scale, largely resolved via DTFs, with residual spectral error clustering in the same low-elevation rear quadrant as the interaural deviations.

This convergence across ITD, ILD, and spectral metrics is consistent with the omission of torso geometry from the synthesis pipeline, since meshes are truncated below the neck. The torso is believed to contribute to HRTFs through two distinct mechanisms: shadowing, which reduces contralateral ear level at low elevations \cite{guldenschuhHRTFModelingDue2008,gumerovNumericalStudyInfluence2002}, and reflection on the shoulders, causing a delayed secondary pulse, which both perturbs onset timing and produces ipsilateral comb filter notches\cite{lanEffectTorsoReflections2022}. Its absence in the synthetic condition therefore plausibly explains the reduced contralateral shadowing that yields the ILD underestimation, while the missing delayed reflection yields both a cleaner apparent onset, producing the ITD overestimation, and the loss of the comb filter interference that would otherwise appear as spectral error in the same region.

KEMAR showed a different deviation profile from measured than synthetic. Its ITD and ILD deviations were significant across a broader proportion of locations, with peak deviations near the horizontal plane rather than at low elevations. The spatial distribution of KEMAR's LSD was similar to synthetic, though concentrated slightly more directly behind the listener. Notably, KEMAR's peak binaural deviations occur near the horizontal plane, the region most commonly used when rendering with KEMAR in prior work \cite{byrneJudgingNumberGender2025, franclDeepNeuralNetwork2022, salorio-corbettoEvaluatingSpatialHearing2022}, where listeners would experience a systematic mismatch from the cues generated by their own anatomy.

\subsection{RQ2: Do computational models predict systematic differences in localisation performance for synthetic HRTFs}

Both auditory models predicted a consistent ordering across HRTF conditions, with measured HRTFs performing best, KEMAR worst by a wide margin, and synthetic and random HRTFs intermediate across front-back confusion, polar error, and polar precision. Only the Barumerli model, which jointly weights ITD, ILD, and spectral features, predicted a modest additional polar cost for synthetic over random, plausibly because Baumgartner relies solely on monaural spectral comparison and cannot penalise the interaural deviations already identified numerically for synthetic HRTFs. In the lateral dimension, via Barumerli, this ordering broke down where KEMAR alone showed a pronounced bias, while measured, random, and synthetic were comparable, though all non-individual conditions were less accurate and precise than measured. Spatially, in the smaller subset of 20 subjects retained for direct comparison with behaviour, predicted errors clustered at high elevations across azimuths for measured and synthetic, likely reflecting the model's directional prior biasing estimates toward the horizon\cite{barumerliBayesianModelHuman2023}, and at the posterior midline (180°) for KEMAR, more plausibly reflecting a direct mismatch between the rendered stimulus and the listener's own spectral templates.

\subsection{RQ3: Do synthetic HRTFs preserve behavioural localisation performance, and does this align with numerical deviations}

Behaviourally, synthetic HRTFs performed as well as measured across every polar metric tested, polar accuracy, precision, and front-back confusion alike, with no significant differences between the two. KEMAR, by contrast, was significantly worse than both on every metric. This closely mirrors the pattern predicted by the computational models, where synthetic and measured were similarly matched and KEMAR incurred by far the largest cost.

Our results are broadly in line with prior work. Using the same localisation test, Daugintis et al. (2026)\cite{daugintisPerceptualEvaluationAuditory2026} reported measured lateral error of $\approx$ 15° and great circle error of $\approx$ 33°, slightly higher than the present measured condition ($\approx$ 9° lateral absolute accuracy, $\approx$ 25° great circle error), with front-back confusion closely matched ($\approx$ 7\% vs. $\approx$ 8\% here). This front-back confusion rate is also comparable to rates reported only after extensive training elsewhere, including $\approx$ 7\% after 1,200 to 1,600 trials \cite{majdak3DLocalizationVirtual2010,middlebrooksIndividualDifferencesExternalear1999}, and lateral and polar errors of a measured control group across multiple sessions (9.3–12.5° lateral, 34–47° polar error) \cite{stittAuditoryAccommodationPoorly2019}. By contrast, single-session, untrained baselines have been reported considerably higher (e.g. 39–43° great circle error, 12–13° lateral error\cite{meyer_generalization_2025}).  Together, these comparisons, while limited by differing metric definitions across studies, support the validity of the present short-form VR-based protocol.

Neither the numerical deviations identified above nor the model's spatial predictions matched where behavioural error occurred. Polar accuracy errors clustered around the front-back midline at low elevations for measured, synthetic, and KEMAR alike, not at the locations where ITD, ILD, and LSD deviations were largest, and the Barumerli model's predicted error showed no significant spatial correlation with behaviour for synthetic on any metric, with a significant negative correlation for KEMAR's polar accuracy. A study using a fully within-subject design across five HRTF conditions offers a useful point of comparison here\cite{pirardEvaluationHeadRelatedTransfer2026}, where a randomly selected HRTF outperformed KEMAR on several polar metrics despite no individualisation at all, consistent with both models predicting random HRTFs to sit closer to measured than KEMAR.

This suggests that factors beyond both the numerical and model metrics are shaping localisation error. Several possibilities exist: the numerical deviations may simply be too small to produce a detectable effect in either measure, front-back confusion may dominate and swamp a smaller structured bias, the numerical analysis covers 200 subjects against just 20 for the behavioural and model comparisons, or again LSD and the positive spectral gradients extracted from gammatone-filtered excitation patterns by the Barumerli model may not capture the perceptual dimensions that actually drive localisation error. Perceptually weighted extensions based on critical bands\cite{bonduLookingRelevantSimilarity2006}, or approaches integrating auditory models of the middle and inner ear, binaural weighting, and perceptual scoring\cite{yaoPerceptuallyEnhancedSpectral2024}, may offer better estimates. Future work will likely require larger datasets combining behavioural localisation data with numerical and modelled predictions across many subjects and a wide range of HRTF types, sufficient to identify what metrics best predict behavioural localisation. 

\subsection{RQ4: Do synthetic HRTFs preserve functional spatial hearing benefits in a speech-in-noise task}
Spatial release from masking increased systematically with masker-target separation in elevation across all HRTF conditions, and neither synthetic nor KEMAR differed significantly from measured. This null result contrasts with prior median-plane SRM research comparing measured and KEMAR HRTFs, which has generally found a significant benefit for individualised HRTFs \cite{gonzalez-toledoSpatialReleaseMasking2024,vicenteExploringRelationshipTask2026}, and there are a few plausible reasons. Overall intelligibility was also considerably lower here than in comparable studies where percentages of correct answers in Vicente et al. (2026) ranged from approximately 60-72\% for individual HRTFs and 53-75\% for KEMAR, compared to roughly 25\% at co-location rising to 45\% at the largest separation in the present study. The target-to-masker ratio in the present study was approximately -3 dB, compared to 0 dB in both prior studies, which may have made the task too difficult to allow detectable differences between HRTFs, consistent with the substantially lower absolute performance observed here. Individual variability also appeared substantial, increasing from 7.6\% at co-location to 13.3\% at the largest separation, and may have further limited sensitivity to a subtle HRTF effect. Native language may also be a factor, as Vicente et al. (2026) found that native English speakers showed no significant benefit of measured over KEMAR, relying instead on spectral glimpsing, with a significant HRTF effect only emerging for non-native listeners or under increased task difficulty. As the present sample was not stratified or modelled by English proficiency, an unaccounted mix of native and non-native speakers could similarly have masked an HRTF effect present in only part of the sample. Additionally, talker gender was a significant factor in the present data, with male talkers detected more accurately than female talkers. Future SRM designs should therefore account for gender imbalance in target-masker pairing, as this could confound apparent spatial or HRTF-related effects.

\section{Conclusion}
Synthetic HRTFs from high-fidelity scans offer a scalable, perceptually valid alternative to acoustic measurement. This study provides the first large-scale numerical and perceptual evaluation of synthetic HRTFs at this scale, combining acoustic analysis across 200 subjects with computational modelling and behavioural testing in smaller cohorts. Synthetic HRTFs numerically approximate measured HRTFs closely, with their residual deviations largely due to the omission of torso geometry from the synthesis pipeline, rather than a general failure of individualisation. Behaviourally, this translates into synthetic HRTFs supporting polar localisation indistinguishable from individually measured HRTFs, while KEMAR does not. At the same time, the spatial signature of the numerical torso bias was not recovered in either behavioural error or the auditory model's own predictions, indicating that current numerical and model-based metrics, while useful for flagging where synthetic HRTFs deviate from measured, do not yet fully explain where or why listeners fail to localise. 

This points to a persistent gap in understanding the relationship between individual ear and torso morphology, the resulting HRTF, and behavioural outcome, which may partly reflect that the acoustic cues a given listener relies on are themselves idiosyncratic and context dependent. Closing this gap will require analytical work that examines these relationships at the level of individual spatial locations and individual listeners, using datasets that pair behavioural outcomes with acoustic and morphological data at sufficient scale to characterise this variability directly.

\section{Supplementary material}
See supplementary material at [URL will be inserted by AIP] for tables of the full generalised linear model outputs of the behavioural analysis for the localisation task and speech task.

\begin{acknowledgments}
This research was supported by the European Union within the SONICOM project (Grant No. 101017743, RIA action of Horizon 2020). I thank Xianghao Wang and Vihan Gemawat for their help in processing the 3D scans to make them suitable for HRTF synthesis.

\end{acknowledgments}

\section{Author declarations}
The authors have no conflicts to disclose. 

\section{Ethics approval}
Experimental procedures were approved by the Imperial College London Ethics Committee (SETREC number: 7046527), and written informed consent was obtained from all participants prior to testing.

\section{Data availability}
The synthetic HRTFs can be found in the Extended SONICOM dataset: \url{https://doi.org/10.61782/fa.2025.0864}. Code to analyse the data can be found in the Spatial Audio Metrics Toolbox here: \url{https://github.com/Katarina-Poole/Spatial-Audio-Metrics}. The localisation behavioural data can be found in the SONICOM Ecosystem here: \url{https://ecosystem.sonicom.eu/databases/58}. Any further data and code used in this study will be made available on reasonable request.

\section{\label{sec:99} References}
\vspace{-10pt}
\bibliography{main}



\end{document}